\documentclass[twocolumn,amsmath,aps]{revtex4}

\usepackage{epsfig}
\usepackage{subfigure}

\newcommand{\be}{\begin{equation}}
\newcommand{\ee}{\end{equation}}
\newcommand{\bea}{\begin{eqnarray}}
\newcommand{\eea}{\end{eqnarray}}

\begin{document}

\title{Living in a Void: Testing the Copernican Principle with Distant Supernovae}
\author{Timothy Clifton} \email{tclifton@astro.ox.ac.uk}
\affiliation{Oxford Astrophysics, Physics, DWB, Keble Road, Oxford, OX1 3RH, UK}
\author{Pedro G. Ferreira}
\affiliation{Oxford Astrophysics, Physics, DWB, Keble Road, Oxford, OX1 3RH, UK}
\author{Kate Land}
\affiliation{Oxford Astrophysics, Physics, DWB, Keble Road, Oxford, OX1 3RH, UK}


\begin{abstract}
A fundamental presupposition of modern cosmology is the Copernican Principle; 
that we are not in a central, or otherwise special region of the Universe.  
Studies of Type Ia supernovae, together with the Copernican Principle,
have led to the inference that the Universe is accelerating
in its expansion.  The usual explanation for this is that there
must exist a `Dark Energy', to drive the acceleration.
Alternatively, it could be the case that the Copernican
Principle is invalid, and that the data has been interpreted
within an inappropriate theoretical frame-work.  If we were to live in
a special place in the Universe, near the centre of a void where the local matter density
is low, then the supernovae observations could be accounted
for without the addition of dark energy. We show that the local redshift dependence of
the luminosity distance  can be used as a clear discriminant between these two paradigms.
Future surveys of Type Ia supernovae that focus on a redshift range of
$\sim 0.1-0.4$ will be ideally
suited to test this hypothesis, and hence to observationally determine
the validity of the Copernican Principle on new scales, as well as probing the degree to which dark 
energy must be considered a necessary ingredient in the Universe.
\end{abstract}

\maketitle



The concordance model of the Universe combines two fundamental
assumptions.  The first is that space-time is dynamical, obeying Einstein's
Equations.  The second is the `Cosmological Principle', that the
Universe is then homogeneous and isotropic on large scales -- 
a generalisation of the Copernican Principle that ``{\it the Earth is
not in a central, specially favored position}'' \cite{Bondi}.  As a result 
of these two assumptions we can use the Freidmann-Robertson-Walker
(FRW) metric to 
describe the geometry of the Universe in terms of a single function, the scale factor
$a(t)$, which obeys
\bea
\label{FRW}
H^2=\frac{8\pi G}{3}\rho-\frac{k}{a^2}
\eea
where $H\equiv \dot a/a$ is the Hubble rate, $\rho$ is the energy
density, $k$ is the (constant) curvature of space, and overdots
denote time derivatives.  The scale factor can then be determined by
observing the `luminosity distance' of astrophysical objects.  At small $z\equiv
a_0/a(t)-1$ this is given by
\be
H_0D_L\simeq c z+\frac{1}{2}(1-q_0)c z^2,
\label{DL}
\ee
where $q\equiv -\ddot a a /\dot{a}^2$ is the deceleration rate, and
subscript $0$ denotes the value of a quantity today. Recent
measurements of ($z$, $D_L$) using high redshift, Type Ia Supernovae (SNe)
have indicated that $q_0<0$, i.e. the Universe is accelerating in its
expansion~\cite{perl, riess}.  Accelerating expansion is possible in
an FRW universe if a fraction of $\rho$ is in the form of
a smoothly distributed and gravitationally repulsive exotic substance,
often referred to as Dark Energy~\cite{albrecht}. The existence of such an
unusual substance is unexpected, and requires previously unimagined amounts of fine-tuning 
in order to reproduce the observations.  Nonetheless, dark energy has been
incorporated into the standard cosmological model, known as $\Lambda$CDM.

\begin{figure*}[htb]
\center
\vspace{-100pt}
\begin{minipage}{15cm}
\subfigure[]{\epsfig{figure=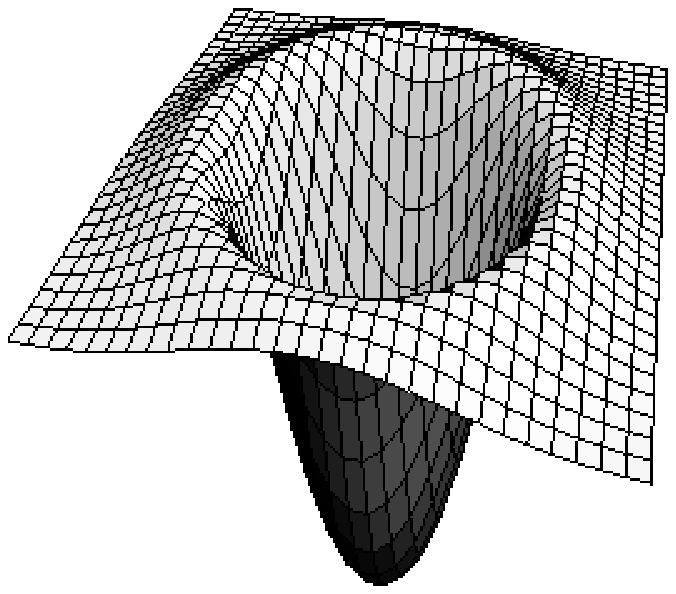,height=3.5cm}\hfill
\; \; \;} \;
\subfigure[]{\epsfig{figure=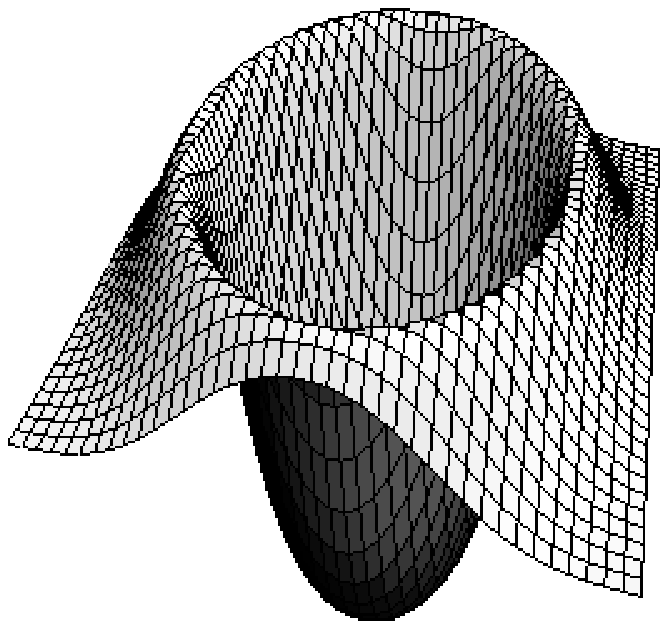,height=3.5cm}\hfill
\; \; \;} \;
\subfigure[]{\epsfig{figure=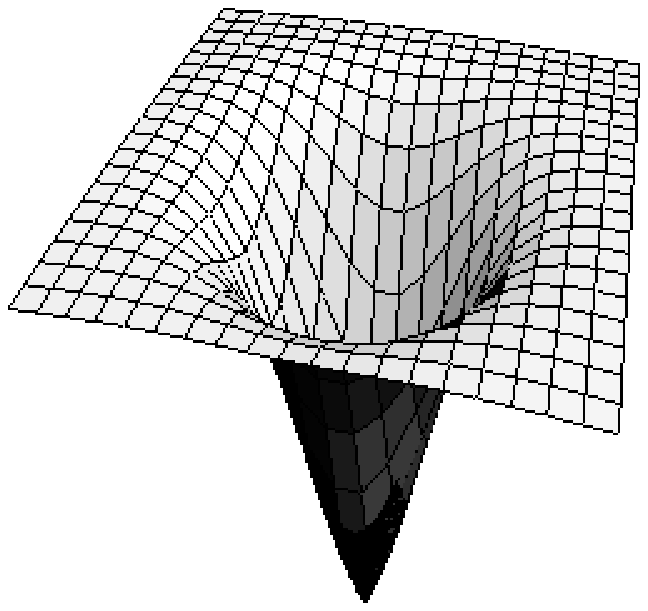,height=3.5cm} \; \; \;}
\vspace{-270pt}
\end{minipage}
\epsfig{figure=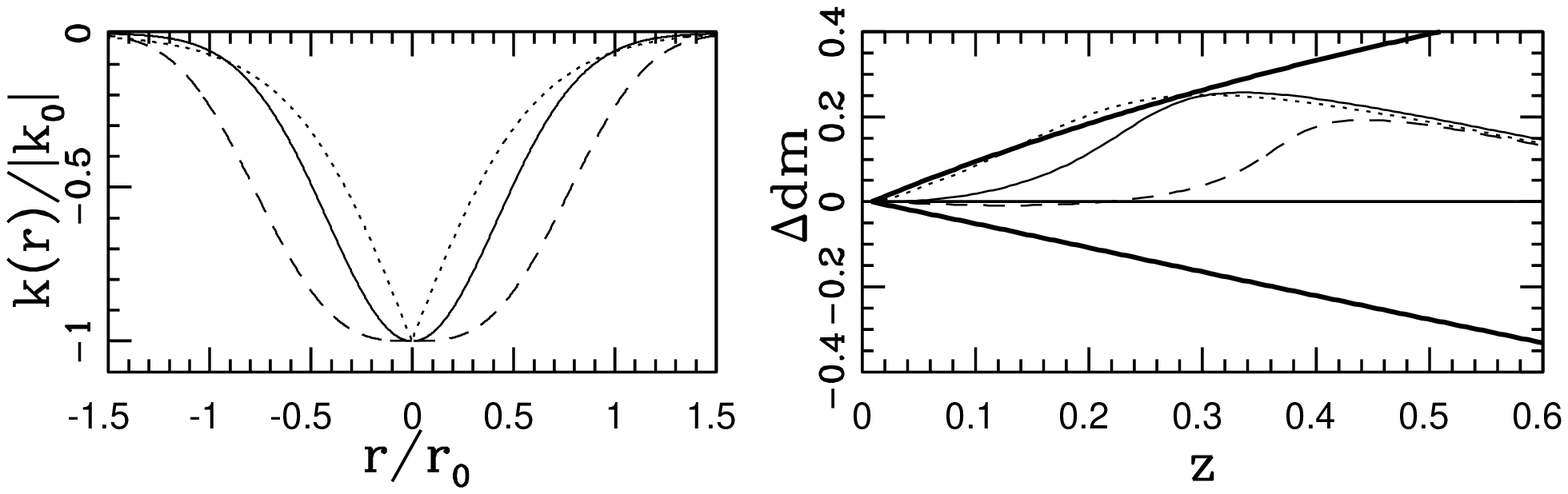,width=15cm}
\vspace{-110pt}
\caption{Some representative forms for the
  curvature function, $k(r)$, and the corresponding distance moduli, 
  normalised to an empty Milne cosmology, $\Delta$dm.  Void (a) is a
  Gaussian in $r$ (with FWHM $r_0$), void (b) is a $k \propto \exp \{
  -c \vert r \vert^3 \} $, and void (c) is $k \propto \{ 1-\vert \tanh (r) \vert \}$, 
  all normalised to the curvature minimum $k_0$.  The three diagrams
  on top indicate the spatial distribution of energy density in the
  three voids at some sample time (solid line for void (a), dashed line
  for void (b) and dotted line for void (c)).
  The ascending thick solid line on the right hand plot corresponds to a universe
  containing dark energy only (de Sitter space), and the descending
  one to a flat universe containing only dust
  (an Einstein-de Sitter universe).}\label{profiles}
\end{figure*}

An alternative to admitting the existence of dark energy is to
review the postulates that necessitate its introduction.  In
particular, it has been proposed that the SNe observations
could be accounted for without dark energy if our local environment were emptier
than the surrounding Universe, i.e. if we were to live in a void~\cite{alexander,alnes,bellido}.  
This explanation for the apparent acceleration does not invoke any exotic substances, 
extra dimensions, or modifications to gravity -- but it does require a rejection of the 
Copernican Principle.  We would be required to live near the centre of
a spherically symmetric under-density, on a scale of the same order of
magnitude as the observable Universe.  Such a situation would have
profound consequences for the
interpretation of all cosmological observations, and would ultimately
mean that we could not infer the properties of the Universe at
large from what we observe locally.

Within the standard inflationary cosmological model the probability of
large, deep voids occurring is extremely small. However, it can be
argued that the centre of a large underdensity is the most likely place for
observers to find themselves \cite{Linde}.  In this case,
finding ourselves in the centre of a giant void would violate
the Copernican principle, that we are not in a special place, but
it may not violate the Principle of Mediocrity, that we are a `typical'
set of observers. Regardless of what we consider the {\it a priori} likelihood of
such structures to be, we find that it should be possible for observers at their centre
to be able to observationally distinguish themselves from their
counterparts in FRW universes.  Living in a void leads
to a distinctive observational signature that, while broadly similar
to $\Lambda$CDM, differs qualitatively in its details.  This gives us
a simple test of a fundamental principle of modern cosmology, as well
as allowing us to subject a possible explanation for the observed
acceleration to experimental scrutiny.

Some efforts have gone into identifying the observational signatures
that could result from living in
a void. The cosmic microwave background (CMB) supplies us with the tight
constraint that we must be
within $15$ Mpc of the center of the void \cite{alnesCMB}. There have
also been some attempts 
at calculating predictions for CMB anisotropies and large
scale-structure \cite{clarksonCMB,zibin,sarkar}, as well as the kinematic
Sunyaev-Zeldovich effect \cite{KSZ}.


General Relativity allows a simple description of
time-dependent, spherical symmetric universes: the
Lema\^itre-Tolman-Bondi (LTB) models \cite{LTB1,LTB2,LTB3}, whose
line-element is
\be
ds^2 = -dt^2 + \frac{a_2^2(t,r) dr^2}{1-k(r) r^2} + a_1^2(t,r) r^2 d \Omega^2,
\ee
where $a_2 = (r a_1)^\prime$, and primes denote
$r$ derivatives .  The old FRW scale factor, $a$,
has now been replaced by two new scale factors, $a_1$ and $a_2$, describing expansion in
the directions tangential and normal to the surfaces of spherical
symmetry.  These new scale factors are functions of time, $t$, and
distance, $r$, from the centre of symmetry, and obey a generalization
of the usual Friedman equation, (\ref{FRW}), such that
\be
\left(\frac{\dot a_1}{a_1}\right)^2 = \frac{8\pi G}{3}{\tilde
  \rho} -\frac{k(r)}{a_1^2}.
\ee
Here ${\tilde\rho} =m(r)/a_1^3$, and is related to the physical energy density by $\rho =
{\tilde \rho}+\tilde{\rho}^\prime r a_1/3 a_2$.  The two free functions, $k(r)$ and $m(r)$, correspond to 
the curvature of space, and the distribution of gravitating mass in that
space.  We choose initial conditions such that the curvature is
asymptotically flat with a negative perturbation near the
origin, and so that the gravitational mass is evenly
distributed.  As the space-time evolves the energy density in the vicinity of
the curvature perturbation is then dispersed, and a void forms.  Observations of distant 
astro-physical objects in this space-time obey a distance-redshift
relation
\be
D_L = (1+z)^2 r_E a_1(t_E,r_E)
\ee
where
\be
1+z = \text{exp} \left\{ \int^{r_E}_{0} \frac{(\dot{a}_1 r)^{\prime}}{\sqrt{1-kr^2}} dr \right\},
\ee
and subscript $E$ denotes the value of a quantity at the moment the observed photon
was emitted.  This expression is modified from equation (\ref{DL}),
allowing for the possibility of apparent
acceleration without dark energy.

We find that the form of the void's curvature profile is of great importance 
for the observations made by astronomers at its centre. 
In Figure~\ref{profiles} we plot some simple curvature profiles,
together with the corresponding
distance moduli as functions of redshift (distance
modulus, $\Delta$dm, is defined as the observable magnitude of an astrophysical
object, minus the magnitude such an object would have at
the same redshift in an empty, homogeneous Milne universe).
It is clear from Figure~\ref{profiles} that for the void models there is a strong
correlation between $k(r)$ and $\Delta$dm; at low redshifts
$\Delta$dm$(z)$ traces the shape of $k(r)$.  Hence, for a generic,
smooth void $\Delta{\rm dm}$ starts off with near
zero slope, where it is locally very similar to a Milne universe,
it then
increases at intermediate $z$, and later drops off like an Einstein-de
Sitter universe. For $\Lambda$CDM, we have $\Delta{\rm
  dm}\simeq-\frac{5}{2}q_0z$ at low $z$, i.e. a non-zero slope. 
Thus, although one can 
always find a void profile that will mimic $\Lambda$CDM~\cite{ellis}, 
such a void will have a curvature profile that is strongly 
cusped, and non-differentiable, at $z=0$~\cite{flan} (see the dotted
line in Figure \ref{profiles} or \cite{yoo}).  Conversely, any
generalized dark energy model capable of producing a flat $\Delta{\rm
  dm}$ at low $z$ would be required to change the equation of
state extremely rapidly between $z\simeq 0.5$ and $0.1$.
\emph{We therefore have a definitive way to
distinguish between a realistic smooth void model, and $\Lambda$CDM}.


\begin{figure}[htbp]
\vspace{-90pt}
\epsfig{figure=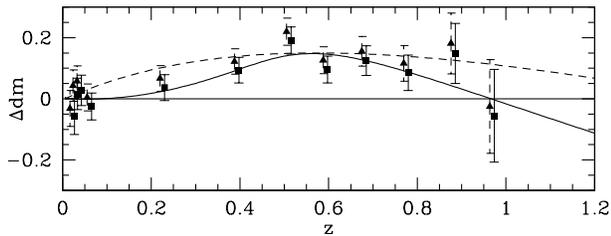,height=8cm}
\vspace{-50pt}
\caption{The current best 
fit $\Lambda$CDM and Gaussian void models as dashed and solid lines,
with triangular and square data points, 
respectively. The data 
shown here is a compilation of 115 low and high-$z$ SNe from
the SNLS, fitted using SALT~\cite{legacy}. For 
illustrative purposes we have binned the results with 10 SNe
per bin (except the last one, which contains 5).  Due to the
uncertainty in the `nuisance' parameters of the calibration 
the data points and the
error-bars shift when fitting for different 
models.}\label{data}\end{figure}

We will now compare the smooth void model to the first-year 
SNe Legacy Survey (SNLS) data, consisting of 115 SNe \cite{legacy} 
calibrated with the 
SALT light-curve fitter\cite{Guy}, and contrast it with
$\Lambda$CDM.  We use the Bayesian information criterion as a figure
of merit (see e.g. \cite{Lid1,Lid2}), and assume one model to be decisively 
favored over the other if
\be
\vert \Delta \ln E \vert \approx \vert \Delta 
(\ln \mathcal{L}_{max}-\frac{p}{2} \ln N) \vert >5,
\label{be}\ee
where $E$ is the evidence for a model, $\mathcal{L}_{max}$ is the maximum likelihood of 
a model, given a data set,  $p$ is the number of parameters in the model, and $N$ is the number of
data points.  This criterion corresponds to one model being $\sim$150 times more likely than the other. 
The minimal void model under consideration has 6 parameters: 2 to parametrize
a Gaussian $k$,
and 4 `nuisance' parameters required to calibrate the
SNe data.  These are absolute magnitude, intrinsic error, and
the colour and stretch parameters used in light curve fitting, 
$\{M_0, \sigma_{\rm int} , \alpha, \beta \}$. 
Assuming spatial flatness, $\Lambda$CDM requires 5 parameters:  1
specifying the fraction of dark energy, and the same 4 nuisance
parameters.  In a more comprehensive study it may be preferable to
perform a full Bayesian evidence analysis, with suitable priors \cite{Lid2}.  In the
interests of brevity, and to avoid a lengthy discussion of prior
probabilities, we have refrained from this for now.

In Figure \ref{data} we show the SNLS data with the two best fit models.  Both have 
similar goodness of
fit, but one can discern a qualitative difference between them, which will be 
distinguishable with future surveys. The best fitting void is $71\pm 7 \%$
underdense at its centre, and has a scale corresponding to $850 \pm 170h^{-1}$Mpc
today.  This is of the order expected to produce a feature in
$\Delta$dm on a scale of $z \sim 0.6$, and large enough to avoid
strong constraints from galaxy surveys that extend to $z\sim
0.1$.  On the other hand,  the best fitting $\Lambda$CDM model 
contains $74 \pm 4\%$  dark energy, and fits the data slightly better 
with $\vert \Delta \ln E \vert \simeq 2.7$.  Thus, while the current
data marginally prefers $\Lambda$CDM, it is not yet able to distinguish between the two models decisively.

One will, of course, be interested in results from
other SNe compilations.  Using the 
Riess gold data~\cite{Riess}, with the MLCS2k2 light-curve fitter, 
we find our basic results do not change significantly, with a $\Lambda$CDM model 
still being marginally preferred.  Thus our analysis does not appear
to be substantially effected by the apparent systematic error that led
to the identification of the `Hubble Bubble' anomaly~\cite{Jha}.  
The `Union' data of ~\cite{SCP} is the largest compilation of SNe
fitted for with the more conservative SALT fitter, and for these 315
SNe (including the `outliers') we find 
the void model is marginally preferred over $\Lambda$CDM, with $\vert
\Delta \ln E \vert \simeq 2.5$ in favor of a void.  It therefore
appears neccessary to obtain more data, in order to be able to
decisively distinguish the two models.


\begin{figure}[htbp]
\vspace{-90pt}
\epsfig{figure=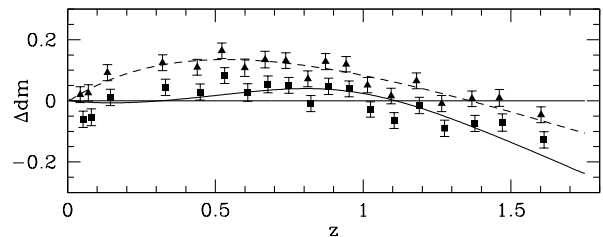,height=8cm}
\vspace{-50pt}
\caption{The best 
fit $\Lambda$CDM and Gaussian void models as dashed and solid lines,
with triangular and square data points, 
respectively, for an example of 700 SNe simulated from a $\Lambda$CDM 
model using the SALT light curve fitter.   The shape of the redshift 
distribution used here is similar to that expected from the 2000 JDEM SNe, with
an extra 300 at low z. We find that 700 SNe, with this redshift
distribution, can decisively recover the 
$\Lambda$CDM model 99\% of the time, as it becomes evident that a 
void model cannot mimic the low redshift behaviour of a $\Lambda$CDM cosmology.
For illustrative purposes, 
we have binned 
in groups of 60.
}
\label{sim}\vspace{-8pt}
\end{figure}

Due to the different strategies and technologies, SNe surveys typically 
target either low \emph{or} high redshifts. This has 
lead to a dearth of SNe at $0.1 \lesssim z \lesssim 0.4$ --
exactly the location where there is the greatest qualitative  
difference between the two models. 
Future SNe surveys, with a redshift coverage in this region, will do better. 
As an example of the future constraints we can expect to gain, we consider the 
JDEM missions which expect to 
observe $\sim$2000 high redshift SNe in the interval $0.1 \lesssim z 
\lesssim 1.7$, with an expected smooth distribution \cite{Kim}. At very low 
redshifts ($0.03 \lesssim z \lesssim 0.1$) a further $\sim$300
SNe can be expected to be observed by other projects. 
We consider data simulated from a $\Lambda$CDM model, using the SALT
fitter, and a Gaussian 
void model with the same parameter values as before.  In both cases 
we find that these 2300 SNe are sufficient to decisively recover the
correct underlying model. For the void model we find $\Delta \ln E=89 \pm 12$, 
while for the $\Lambda$CDM model $\Delta \ln E=46 \pm 10$.
These are considerably better 
than the decisive benchmark, which was satisfied by all 1000
of our simulations (except one in the $\Lambda$CDM case). We can therefore 
say with confidence that the upcoming SNe data will be able to distinguish $\Lambda$CDM 
from a void model, independent of which is responsible for the apparent acceleration. 

Given that $\sim$2300 SNe from a JDEM mission will do a superfluous job, 
we will now estimate the minimum number of SNe required to meet the
bound in Equation (\ref{be}). 
Using the same redshift distribution as before \cite{Kim}, we now consider 
different numbers of SNe. In the case of the void cosmology
we find that with $\sim$170 SNe 50\% of our 1000 
simulations recovered the void 
model decisively, while with $\sim$480 SNe 99\% of the
simulations could do so. 
Similarly, in the case of a $\Lambda$CDM cosmology  we found that with 
$\sim$180 and $\sim$700 SNe we could decisively 
recover the correct model in 50\% and 99\% of the simulations, respectively. This is illustrated in Fig. 3. These 
projections are much lower 
than the number of SNe expected to be observed by the next
generation of SNe surveys, 
and we should therefore expect less ambitious 
projects to able to distinguish between the two models.  In fact, this
may be possible soon, with data from the Sloan Survey
and third year SNLS expected to be released imminently.  The results
obtained here depend on the details of the future surveys.
Tailoring these to the specific regions where the two models differ
most would undoubtedly give decisive results sooner.

Consider now the effect of varying the redshift distribution, rather
than the overall number.  Adding a further 300 SNe to the SNLS data,
and letting them have a Gaussian redshift distribution 
with $\sigma_z=0.1$ and a mean that can vary, we simulate the data from void and 
$\Lambda$CDM models. We find that 
in a void Universe the optimal place to search for these extra 300 SNe is at
$z\sim 0.3$  (limiting ourselves to a mean redshift of less than 1),
where the average  $\Delta \ln {\mathcal L}_{\rm max} \sim 5$. In a  
$\Lambda$CDM universe the SNe would be better placed a little lower,
at $z\sim 0.1$, and in this case $\Delta \ln {\mathcal L}_{\rm max} \sim 2.5$.



We emphasise that similar results to those presented above should be obtainable for \emph{any}
smooth void model.  For example, repeating our analysis for void (b),
in Fig. \ref{profiles}, gives almost identical results to void (a).
The reason for this is that all smooth voids display the same qualitative behaviour of having a 
flat $\Delta$dm($z$) profile at low-$z$.  It is this qualitative
difference that allows the void models to be so easily distinguished from
$\Lambda$CDM, and as all smooth voids display this low-$z$ behaviour,
we expect the results we have presented to be broadly generalizable to
all voids.  Of course, it may be possible to imagine anomolous cases.
This will be considered further in a more extended future publication.

Two very different paradigms have been invoked to explain the current
observation of an apparently accelerating Universe,  
depending on whether we invoke or reject the Copernican Principle. We have shown 
that in the coming years it will be possible to experimentally distinguish between these two 
scenarios, allowing us to experimentally test the 
Copernican Principle \cite{uzan,clarkson, caldwell}, as well as
determine the extent to which Dark Energy must be considered a 
neccessary ingredient in the Universe.

\vspace{-20pt}
\section*{Acknowledgements}
\vspace{-10pt}

We are grateful to C. Clarkson, L. Miller, A. Slosar, and M. 
Sullivan for discussions, and the BIPAC for support.  TC acknowledges the support of Jesus College, 
and KL the Glasstone foundation and Christ Church.


\end{document}